\documentclass{jpsj-suppl}
\usepackage{graphicx} 

\newcommand{\be}{\begin{equation}}
\newcommand{\ee}{\end{equation}}
\newcommand{\ba}{\begin{eqnarray}}
\newcommand{\ea}{\end{eqnarray}}
\newcommand{\nn}{\nonumber}
\newcommand{\ci}[1]{\cite{#1}}
\newcommand{\gev}{\,{\rm GeV}}

\newcommand{\tw}{\textwidth}
\newcommand{\req}[1]{(\ref{#1})}
\newcommand{\lsim}{\raisebox{-4pt}{$\,\stackrel{\textstyle <}{\sim}\,$}}
\newcommand{\gsim}{\raisebox{-4pt}{$\,\stackrel{\textstyle
      >}{\sim}\,$}}

\title{The Exclusive Drell-Yan Process and Deeply Virtual Pion Production}

\author{Peter \textsc{Kroll}}

\inst{Fachbereich Physik, Universit\"at Wuppertal, D-42096 Wuppertal, Germany and  \\
Institut f\"ur Theoretische Physik, Universit\"at Regensburg, D-93040 Regensburg, 
Germany}

\email{pkroll@uni-wuppertal.de}

\abst{In this talk it is reported on analyses of $p\to l\pi^+n$ and $\pi^-p\to l^+l^- n$ 
within the handbag approach. It is argued that recent measurements of hard pion 
production performed by HERMES and CLAS clearly indicate the occurrence of strong 
contributions from transversely polarized photons. The $\gamma^{\,*}_T\to \pi$ transitions 
are described by the transversity GPDs accompanied by twist-3 pion wave 
functions. The experiments also require strong contributions from the pion
pole which can be modeled as classical one-pion exchange. With these extensions the 
handbag approach leads to results on cross sections and spin 
asymmetries in fair agreement with experiment. This approach is also used 
for an estimate of the partial cross sections for the exclusive Drell-Yan process.}

\kword{Pion production, factorization, twist-3, transversity, Drell-Yan process}

\begin{document}
\maketitle

\section{Introduction}
\label{sec:intro}
The handbag approach offers a partonic description of many hard exclusive processes.
It is based on factorization in hard subprocesses and soft hadronic matrix elements,
the so-called generalized parton distributions (GPDs). This approach has been applied
by now to many deeply virtual and wide-angle processes and even to the exclusive
production of hardons involving heavy quarks. Here in this talk, the interest is
focused on exclusive pion leptoproduction, $lp\to l\pi^+n$, and the exclusive 
pion-induced Drell-Yan process, $\pi^- p\to l^+ l^-n$. 

For pion leptoproduction there is a rigorous proof \ci{collins96} of collinear 
factorization of the process amplitudes for longitudinally polarized virtual photons, 
$\gamma_L^*$, in a hard partonic subprocess and GPDs in the generalized Bjorken regime 
of large photon virtuality, $Q^2$, and large photon-proton center of mass energy, $W$, 
but small invariant momentum transfer, $-t$. It however turned out that leading-twist 
calculations underestimate the experimental pion production cross section by order of 
magnitude. It became evident as I am going to demonstrate in Sect.\ \ref{sec:lepto},  
the pion pole is to be treated as a classical one-particle exchange rather then as
a part of the GPD $\widetilde E$. Moreover, experiments tell us that there are
strong, for $\pi^0$ production even dominant, contribution from transversely
polarized photons, $\gamma^*_T$.

The high-energy pion beam at J-PARC offers the possibility of measuring the exclusive
Drell-Yan process which is closely related to pion leptoproduction. The same GPDs
contribute and the subprocesses are $\hat s - \hat u$ crossed ones
($\hat s$ and $\hat u$ are the subprocess Mandelstam variables):
\be
{\cal H}^{\pi^-\to\gamma^*}(\hat u, \hat s) \,=\, 
                               - {\cal H}^{\gamma^*\to\pi^+}(\hat s, \hat u)\,.
\label{eq:crossing}
\ee
Eq.\ \req{eq:crossing} is equivalent to the replacement of the space-like
photon virtuality, $Q^2$, by minus the time-like one, $Q^{\prime 2}$. Since 
leading-twist calculations of pion production underestimates experiment by far
it is plausible to expect a similar failure for the Drell-Yan process. As 
advocated for in \ci{GK9}, one may rather apply that what has been learned 
in the analysis of pion production also to the Drell-Yan process. In Sect.\ 
\ref{sec:DY} it is reported on such an analysis. Future data on the 
exclusive Drell-Yan process may reveal whether or not our present understanding
of hard exclusive process in terms of convolutions of GPDs and hard scatterings
also holds for time-like photons. This is a non-trivial issue because the 
physics in the time-like region is complicated and often not well understood.

\section{Leptoproduction of pions}
\label{sec:lepto}
The leading-twist helicity amplitudes for pion production read 
\ba
{\cal M}_{0+,0+} &=& \frac{e_0}{Q}\sqrt{1-\xi^2} \int^1_{-1} dx {\cal H}_{0+0+}  
\Big[\widetilde H  - \frac{\xi^2}{1-\xi^2}  \widetilde E \ \Big]\,, \nn\\ 
{\cal M}_{0-,0+} &=& \frac{e_0}{Q} \frac{\sqrt{-t'}}{2m} \xi 
                                             \int_{-1}^1dx {\cal H}_{0+0+} \widetilde E  \,, 
\label{eq:twist2}
\ea
where ${\cal H}_{0+0+}$ denotes the subprocess amplitude and ${\widetilde H}$, $\widetilde E$ the
relevant GPDs. The nucleon mass is denoted by $m$. The skewness, $\xi$, is related to 
Bjorken-x by $\xi =x_B/(1-x_B)$ up to corrections of order $1/Q^2$. In 
\req{eq:twist2} the usual abbreviation $t'=t-t_0$ is employed where the minimal 
value of $-t$ corresponding to forward scattering, is $t_0=-4m^2\xi^2/(1-\xi^2)$.
Helicities are labeled by their signs or by zero; they appear in the familiar 
order: pion, outgoing nucleon, photon, ingoing nucleon.

Power corrections to the leading-twist result \req{eq:twist2} are theoretically 
not under control. It is therefore not clear at which values of $Q^2$ and $W$ 
the amplitudes \req{eq:twist2} can be applied. The onset of the leading-twist
dominance has to be found out by comparison with experiment. An example of power 
corrections is set by the $\gamma^*_T\to \pi$ amplitudes which are theoretically 
suppressed by $1/Q$ as compared to the $\gamma_L^*\to\pi$ amplitudes. Still, as 
experiments tell us, the $\gamma_T^*\to\pi$ amplitudes play 
an important role  in hard exclusive pion leptoproduction. The first experimental 
evidence for strong contribution from $\gamma_T^*\to\pi$ transitions came from the 
spin asymmetry, $A_{UT}$, measured with a transversely polarized target by the HERMES 
collaboration for $\pi^+$ production \ci{hermes-aut}. Its $\sin{\phi_s}$ 
modulation where the angle $\phi_s$ defines the orientation of the target spin vector,
unveils a particularly striking behavior: It is rather large and does not show any 
indication of a turnover towards zero for $t'\to 0$. For $t'\to 0$ 
$A_{UT}^{\sin \phi_s}$ is under control of the interference term  
\be
 A_{UT}^{\sin \phi_s}\propto {\rm Im} \Big[{\cal M}^*_{0-,++}{\cal M}_{0+,0+}\Big]\,.
\ee
Both the amplitudes are not forced by angular momentum conservation to vanish in 
the limit $t'\to 0$. Hence, the small $-t'$ behavior of the $A_{UT}^{\sin \phi_s}$ data 
entails a sizeable $\gamma_T^*\to\pi$ amplitude.
  
\begin{figure*}
\centering
\includegraphics[width=0.38\tw]{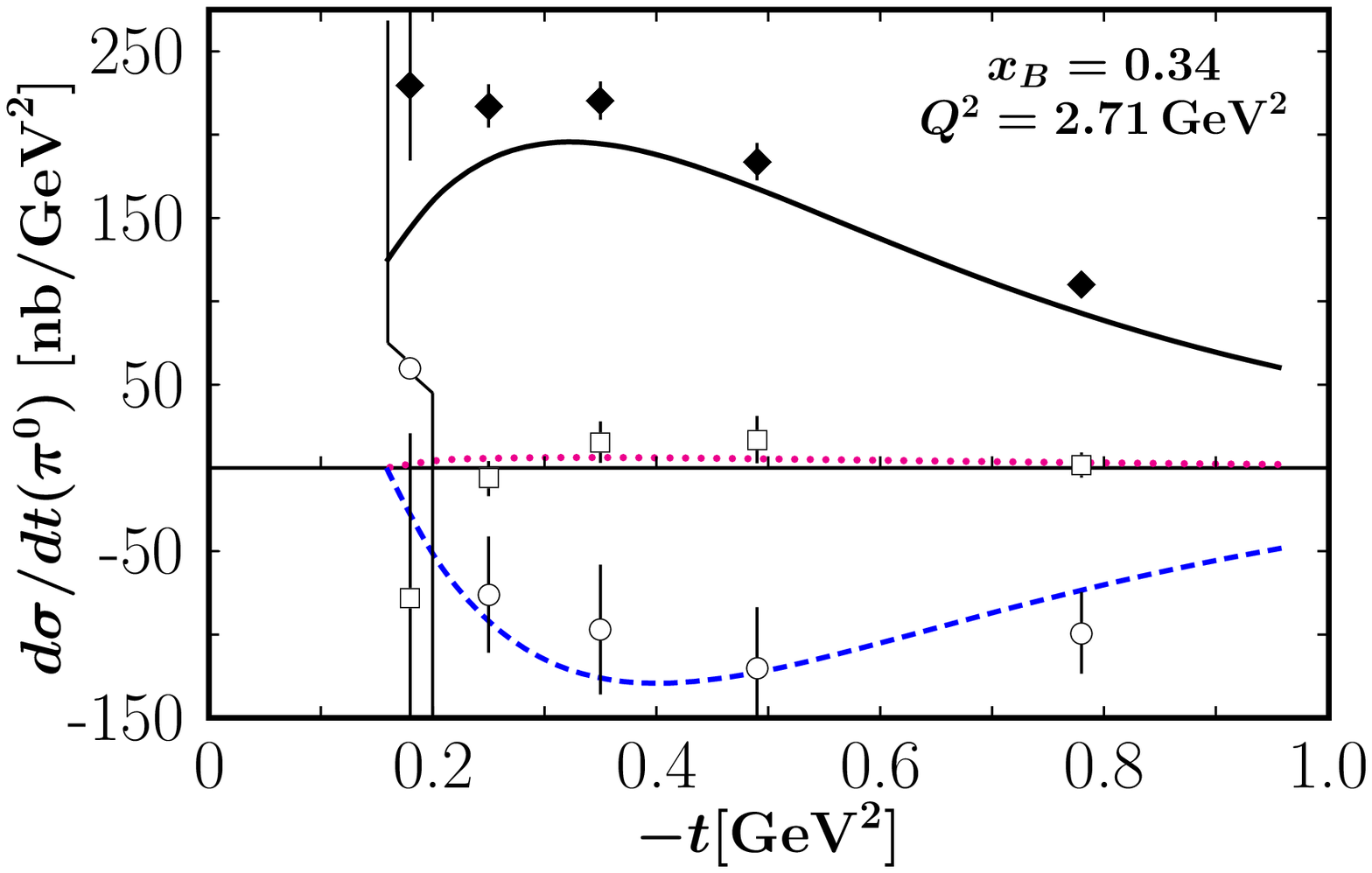}\hspace*{0.05\tw}
\includegraphics[width=0.46\tw]{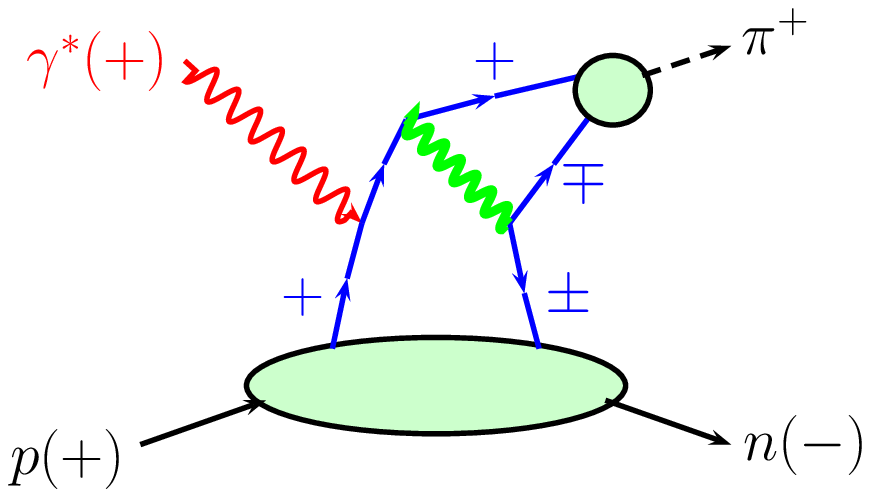}
\caption{\label{fig:1} The unseparated (long.-transv., transv.-transv.) cross 
section. Data \ci{clas-pi0} are shown as diamonds (squares, circles). 
The theoretical results are taken from \ci{GK6}.}
\caption{\label{fig:2} A typical lowest-order Feynman graph for pion 
leptoproduction. The signs indicate the helicities of the involved particles. }      
\end{figure*}

A second evidence for large contributions from transversely polarized 
photons comes from the CLAS measurement \ci{clas-pi0} of the $\pi^0$ 
electroproduction cross sections. As can be seen from Fig.\ \ref{fig:1} the 
transverse-transverse interference cross section is negative and amounts to about
$50\%$ of the unseparated cross section in absolute value. 

%
\subsection{Transversity}
\label{sec:transversity}
The question is whether or not the handbag approach can be generalized in order
to account also the $\gamma^*_T\to \pi$ amplitudes.
In Fig.\ \ref{fig:2} a typical Feynman graph for pion production is depicted
with helicity labels for the amplitude ${\cal M}_{0-,++}$. Angular momentum
conservation forces the subprocess amplitude as well as the nucleon-parton matrix 
element to vanish $\propto \sqrt{-t'}$ for $t'\to 0$ for any contribution to 
${\cal M}_{0-,++}$ from the usual helicity non-flip GPDs $\widetilde H$ and 
$\widetilde E$. This  result is in conflict with the HERMES data on $A_{UT}^{\sin\phi_s}$. 
In \ci{GK5,GK6} (see also \ci{liuti15}), it has been suggested that contributions 
from the transversity GPDs, for which the emitted and 
reabsorbed partons have opposite helicities, are responsible for the above mentioned 
experimental phenomena. Factorization of the $\gamma^*_T\to\pi$ amplitudes is assumed. 
The dominant contributions from the transversity GPDs are ($\bar{E}_T=2\widetilde{H}_T+E_T$)
\ba
{\cal M}_{0+,\mu +} &=& e_0 \frac{\sqrt{-t^\prime}}{4m} \int^1_{-1} dx 
                                             {\cal H}_{0-,++}\bar{E}_T\,, \nn\\
{\cal M}_{0-,++} &=& e_0 \sqrt{1-\xi^2} \int^1_{-1} dx {\cal H}_{0-,++}H_T\,,  \nn\\  
{\cal M}_{0-,-+}&=& 0\,.
\label{eq:twist3-sim}
\ea    
The neglect of the other transversity GPDs is justified at least at small $\xi$ and 
$-t^\prime$. As we will see these amplitudes meet the main features of the experimental 
data mentioned above.

\subsection{The subprocess amplitude ${\cal H}_{0-,++}$}
\label{sec:subprocess}
As can be seen from Fig.\ \ref{fig:2} quark and antiquark forming the pion have the same 
helicity for ${\cal H}_{0-,++}$. Hence, a twist-3 pion wave function is required. There 
are two twist-3 distribution amplitudes, a pseudoscalar one, $\Phi_P$ and a pseudotensor 
one, $\Phi_\sigma$. Assuming the three-particle contributions to be strictly zero, the twist-3 
distribution amplitudes are given by \ci{braun90}
\be
\Phi_P\equiv 1    \qquad \qquad \Phi_\sigma=6\tau (1-\tau)
\label{eq:PDA}
\ee
Here $\tau$ is the momentum fraction the quark in 
the pion carries with respect to the pion momentum. The subprocess amplitude 
${\cal H}_{0-,++}$ is computed to lowest-order of perturbation theory. It turns out 
that the pseudotensor contribution is proportional to $t/Q^2$ and consequently neglected. 
The pseudoscalar term is non-zero at $t=0$ but infrared singular. In order to regularize 
this singularity the modified perturbative approach is used in \ci{GK5,GK6} in which 
quark transverse momentum, $k_\perp$, are retained in the subprocess whereas the 
emission and reabsorption of partons from the nucleons is assumed to happen 
collinear to the nucleon momenta \ci{GK3}. In this approach the subprocess 
amplitude for $\pi^+$ production reads  
\ba
{\cal H}_{0-,++} &=&\frac{2}{\pi^2} \frac{C_F}{\sqrt{2N_C}}\mu_\pi 
        \int d\tau d^2{\bf k}_\perp \Psi_{\pi P}(\tau, k_\perp) \alpha_s(\mu_R) \nn\\
    \hspace{-0.1\tw}  &\times & \Big[\frac{e_u}{x-\xi+i\epsilon}
                     \frac1{\bar{\tau}\frac{x-\xi}{2\xi}Q^2-k_\perp^2+i\epsilon} 
        - \frac{e_d}{x+\xi-i\epsilon}
                     \frac1{-\tau\frac{x+\xi}{2\xi}Q^2-k_\perp^2+i\epsilon}\Big]\,.
\label{eq:twist3-sub}
\ea
For the case of the $\pi^0$ the quark charges have to be taken out; they appear in the 
flavor combination of the GPDs. The pion light-cone wave function, $\Psi_{\pi P}$, is 
parametrized as
\be
\Psi_{\pi P}\,=\,\frac{16\pi^{3/2}}{\sqrt{2N_C}} f_\pi a_P^3|{\bf k}_\perp|
                                                    \exp{[-a_P^2k_\perp^2]}\,.
\ee
Its associated distribution amplitude is the pseudoscalar one given in \req{eq:PDA}. 
For the transverse-size parameter, $a_P$, the value $1.8\,\gev$ is adopted and
$f_\pi (=0.132\,\gev)$ denotes the pion decay constant. The parameter $\mu_\pi$ 
in \req{eq:twist3-sub} is related to the chiral condensate
\be
\mu_\pi \,=\, \frac{m_\pi^2}{m_u+m_d}
\ee
where $m_\pi$ is the pion mass and $m_u, m_d$ current quark masses. 
In \ci{GK5,GK6} a value of $2\,\gev^2$ is taken for $\mu_\pi$. The contributions from 
transversely polarized photons are parametrically suppressed by $\mu_\pi/Q$ as compared 
to the $\gamma^*_L\to\pi$ amplitudes \req{eq:twist2}. However, for values of $Q$ 
accessible in current experiments, $\mu_\pi/Q$ is of order 1. 

The amplitude \req{eq:twist3-sub} is Fourier transformed to the impact parameter 
space and the integrand is multiplied by a Sudakov factor, $\exp{[-S]}$, which 
represents gluon radiation in next-to-leading-log approximation using resummation 
techniques and having recourse to the renormalization group \ci{li-sterman}. The 
Sudakov factor cuts off the $b$-integration at $b_0=1/\Lambda_{\rm QCD}$. In the modified 
perturbative approach the renormalization and factorization scales are 
$\mu_R={\rm max}[\tau Q, (1-\tau)Q,1/b]$ and $\mu_F=1/b$, respectively. In 
\ci{GK5,GK6} the modified appoach is also applied to the $\gamma^*_L\to\pi$ amplitudes. 
The Sudakov factor guarantees the emergence of the twist-2 result for $Q^2\to\infty$.

\subsection{The pion pole}
\label{sec:pion-pole}
\begin{figure*}
\centering
\includegraphics[width=0.46\tw]{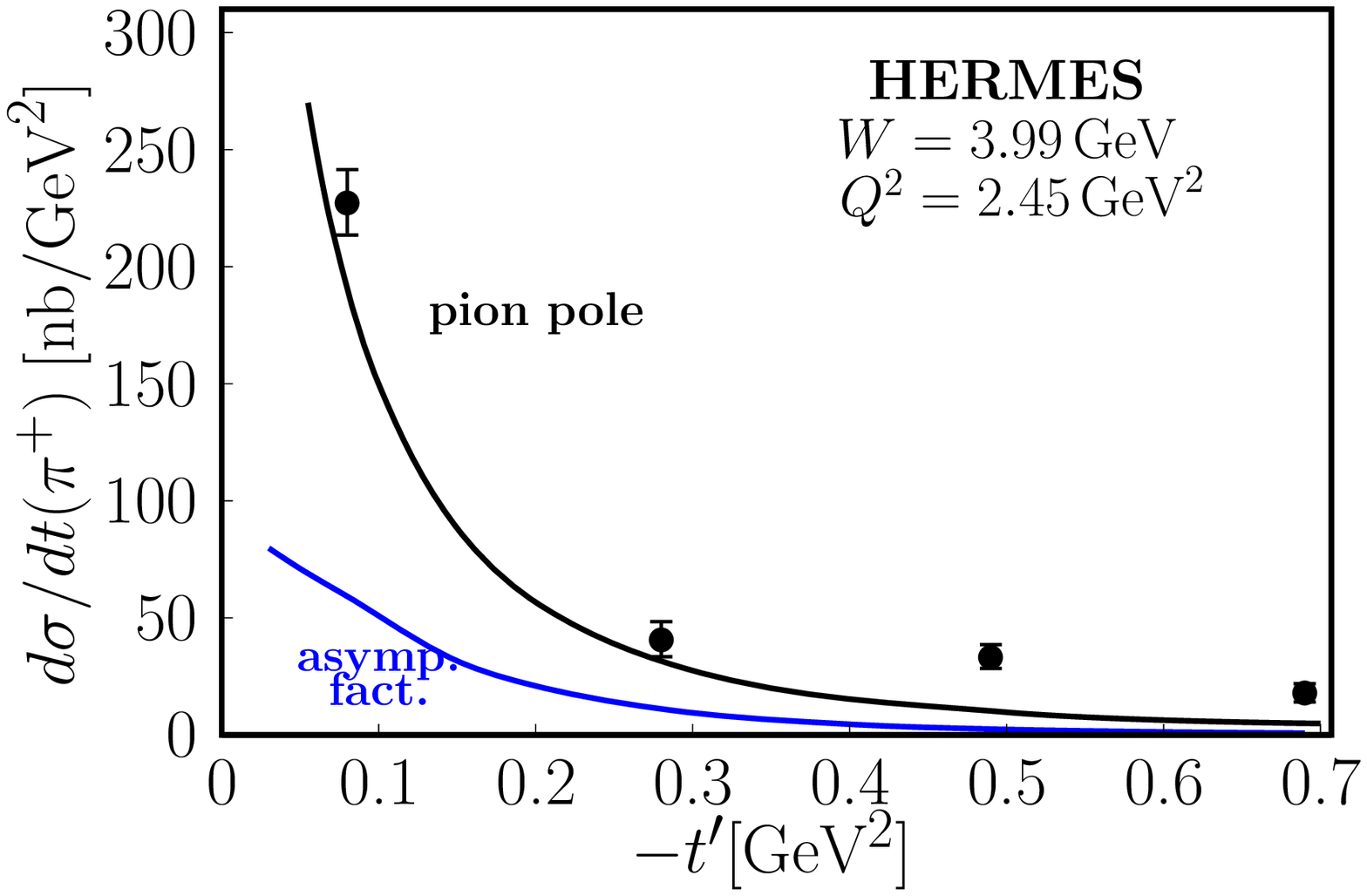}\hspace*{0.05\tw}
\includegraphics[width=0.38\tw]{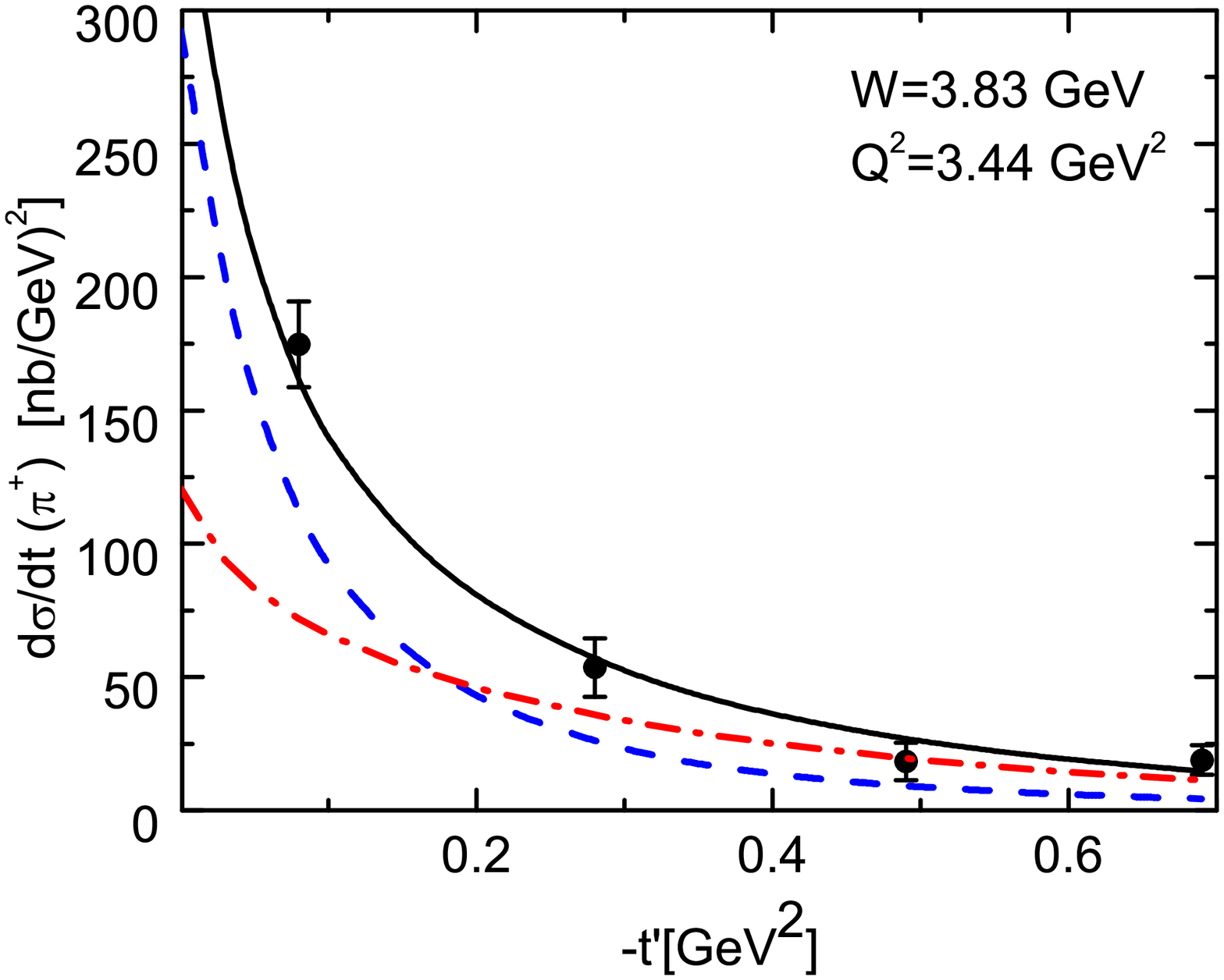}
\caption{\label{fig:3} The unseparated $\pi^+$ cross section versus $-t'$.
The lines represent the pion-pole contribution \req{eq:pole-contr} with 
$F_\pi^{\rm pert}$ replaced by $F_\pi^{\rm exp}$ and the leading-twist result. 
Data taken from \ci{hermes08}. }
\caption{\label{fig:4} The unseparated $\pi^+$ cross section. Data taken from
\ci{hermes08}. The solid (dashed, dash-dotted) curve represents the results of 
\ci{GK5} for the unseparated (longitudinal, transverse) cross section.} 
\end{figure*}

A special feature of $\pi^+$ production is the appearance of the pion pole. As has 
been shown in \ci{mankiewicz99} the pion pole is part of the GPD $\widetilde E$
\be
{\widetilde E}^u_{\rm pole}\,=\,-{\widetilde E}^d_{\rm pole}\,=\,- \Theta(\xi -|x|)
\frac{m f_\pi g_{\pi NN}}{\sqrt{2}\xi}\frac{F_{\pi NN}(t)}{t-m_\pi^2}
                         \Phi_\pi\big(\frac{x+\xi}{2\xi}\big)
\ee
The coupling of the exchanged pion to the nucleons is described by the coupling 
constant $g_{\pi NN} (=13.1\pm 0.2)$ and a form factor parametrized as a monopole with
a parameter $\Lambda_N (=0.44\pm 0.07$); $F_\pi$ denotes the electromagnetic form factor
of the pion. The convolution of ${\widetilde E}_{\rm pole}$ with the subprocess amplitude 
${\cal H}_{0\lambda,0\lambda}$ can be worked out analytically. The contribution of the pion
pole to the longitudinal cross section is
\be
\frac{d\sigma_L^{\rm pole}}{dt}\,\sim\, \frac{-t\, Q^2}{(t-m_\pi^2)^2}
                     \Big[e_0 g_{\pi NN} F_{\pi NN}(t) F_\pi^{\rm pert}(Q^2) \Big]^2\,.  
\label{eq:pole-contr}
\ee
Here, $F_\pi^{\rm pert}$ is the leading-order perturbative contribution to $F_\pi$ which is 
known to underestimate the experimental form factor \ci{F-pi-08} substantially, and 
consequently \req{eq:pole-contr} is much smaller than data on the $\pi^+$ cross section, 
see Fig.\ \ref{fig:3}. In \ci{GK5,GK6} the perturbative result for $F_\pi$ was therefore 
replaced by its experimental value. This prescription is equivalent to computing the pion 
pole contribution as a one-particle exchange. With this replacement one sees that the pole 
term controls the $\pi^+$ cross section at small $-t'$, see Fig.\ \ref{fig:3}. A detailed 
discussion of the pion pole contribution can be found in \ci{FGHK15}. It also plays a 
striking role in electroproduction of $\omega$ mesons \ci{GK8} where it dominantly 
contributes to the amplitudes for transversely polarized $\omega$ mesons.  
%

\subsection{Phenomenology}
\label{sec:phenomenology}
In order to make predictions and comparison with experiment the GPDs are needed.
In \ci{GK5,GK6,GK3} the GPDs are constructed with the help of the 
double-distribution representation. According to \ci{musatov00} a double 
distribution is parametrized as a product of a zero-skewness GPD and an 
appropriate weight function that generates the skewness dependence. The 
zero-skewness GPD itself is composed of its forward limit, $K(x,\xi=t=0)=k(x)$, 
multiplied by an exponential in $t$ with a profile function, $f(x)$, parametrized 
in a Regge-like manner
\be
f(x)\,=\,-\alpha'\ln{x} + B
\ee
at small $-t$. The GPD $\widetilde H$ at $\xi=0$ is taken from a recent analysis
of the nucleon form factors \ci{DK13} while $\widetilde E$ is neglected. The forward 
limit of the GPD $H_T$ is given by the transversity parton 
distributions, known from an analysis of the azimuthal asymmetry in semi-inclusive 
deep inelastic lepton-nucleon scattering and inclusive two-hadron production in 
electron-positron annihilations \ci{anselmino08}, but
the normalization of $H_T$ is adjusted to lattice QCD moments \ci{gockeler05}.
The forward limit of $\bar{E}_T$ is parametrized like the usual parton densities
\be
\bar{e}_T(x)\,=\,N x^{-\alpha}(1-x)^\beta
\ee
with $\alpha=0.3$ for both $u$ and $d$ quarks and $\beta^u=4$, $\beta^d=5$.
The normalization as well as the parameters $\alpha'$ and $B$ for each of the 
transversity GPDs are estimated by fits to the HERMES data on the $\pi^+$ cross 
section \ci{hermes08} and to lattice QCD results \ci{gockeler07} on the 
moments of $\bar{E}_T$.

An example of the results for the $\pi^+$ cross section is shown in Fig.\ 
\ref{fig:4}, typical results for $\pi^0$ production in Fig.\ \ref{fig:1}. At small 
$-t'$ the $\pi^+$ cross section is under control of the pion pole as discussed in 
Sect.\ \ref{sec:DY}. The contribution from $\widetilde H$ to the longitudinal 
cross section is minor. The transverse cross section, although parametrically suppressed by 
$\mu_\pi^2/Q^2$, is rather large and even dominates for $-t'\gsim 0.2\,\gev^2$.
It is governed by $H_T$, the contribution from $\bar{E}_T$ is very small. This
fact can easily be understood considering the relative sign of $u$ and $d$
quark GPDs. For $H_T$ they have opposite signs but the same sign for $\bar{E}_T$.
Moreover, $\bar{E}^u_T$ and  $\bar{E}^d_T$ are of similar size. This pattern of these
GPDs is supported by large-$N_c$ considerations \ci{weiss16}.
Since the GPDs contribute to $\pi^+$ production in the flavor combination
\be
K^{\pi^+} \,=\, K^u - K^d
\ee
it is obvious that the contributions from $\bar{E}_T^u$ and $\bar{E}_T^d$ cancel 
each other to a large extent in contrast to those from $H_T$.    

For $\pi^0$ production the situation is reversed since the GPDs appear now
in the flavor combination
\be
K^{\pi^0}\,=\,\frac1{\sqrt{2}} \big(e_u K^u - e_d K^d\big)\,.
\label{eq:pi0-flavor}
\ee 
The contribution from $\bar{E}_T$ makes up the transverse-transverse interference cross 
section whereas the transverse cross section receives contribution from both $H_T$ and  
$\bar{E}_T$ (see Eq.\ \req{eq:twist3-sim}):
\be
\frac{d\sigma_T}{dt} \,=\,\frac1{\kappa}\Big[\frac12|{\cal M}_{0-,++}|^2
             + |{\cal M}_{0+,++}|^2\Big]   
\qquad \qquad \frac{d\sigma_{TT}}{dt}\,=\,-\frac1{\kappa}|{\cal M}_{0+,++}|^2
\ee
where $\kappa$ is a phase-space factor.
However, the sum of these cross section is only fed by $H_T$. In \ci{GK5,GK6} it is 
predicted that for $\pi^0$ production  
\be
\frac{d\sigma_L}{dt} \ll  \frac{d\sigma_T}{dt}\,.
\label{eq:LT}
\ee
Hence, to a good approximation the transverse cross section equals the unseparated 
one. The prediction \req{eq:LT} is consistent with the very small 
longitudinal-transverse interference cross section, see Fig.\ \ref{fig:1},
and is confirmed by a recent Hall A measurement \ci{hall-A}. It is to 
be emphasized that this prediction is what is expected in the limit $Q^2\to 0$ and 
not for $Q^2\to \infty$. 

The transversity GPDs play a similarly prominent role in leptoproduction of other
pseudoscalar mesons. Consider, for example, $\eta$ production. Under 
the plausible assumption $K^s=K^{\bar{s}}$ only the $u$ and $d$-quark GPDs in the combination
\be
K^\eta \simeq \frac1{\sqrt{6}}\big(e_u K^u + e_d K^d\big)\,.
\label{eq:eta-flavor}
\ee
contribute to $\eta$ production. With regard to the different signs in \req{eq:eta-flavor} 
and \req{eq:pi0-flavor} it is evident that $H_T$ plays a more important role for $\eta$ 
than for $\pi^0$ production whereas for $\bar{E}_T$ the situation is reversed with the 
consequence of an $\eta/\pi^0$ ratio of about 1 for $t'\to 0$ and a small ratio otherwise. 
The transverse-transverse cross section is now very small because of the strong 
cancellation between $\bar{E}_T^u$ and $\bar{E}_T^d$. These properties of $\eta$ production
is in fair agreement with preliminary CLAS data \ci{kubarowsky10}.

\section{The exclusive Drell-Yan process}
\label{sec:DY}
Let me now turn to the Drell-Yan process. Leading-twist analyses have
been performed in \ci{berger,sawada}. It has been found that the longitudinal cross
section amounts to a few ${\rm pb}/\gev^2$ at small $t'$. With regard to the
failure of leading-twist analyses of pion leptoproduction a reanalysis of the Drell-Yan 
process seems to be appropriate taking into account what has been learned from analyses of
pion leptoproduction \ci{GK9}. 

The cross section for the Drell-Yan processs reads
\ba
\frac{d\sigma}{dt dQ^{\prime 2} d\cos{\theta}d\phi} &=&\frac{3}{8\pi}\left\{
            \sin^2{\theta}\,\frac{d\sigma_L}{dtdQ^{\prime 2}}
+\frac{1+\cos^2{\theta}}{2}\, \frac{d\sigma_T}{dtdQ^{\prime 2}} \right.\nn\\
 &+&\left.\frac{\sin{(2\theta)}\cos{\phi}}{\sqrt{2}} \,
               \frac{d\sigma_{LT}}{dtdQ^{\prime 2}}
     + \sin^2{\theta}\cos{(2\phi)}\,\frac{d\sigma_{TT}}{dtdQ^{\prime 2}}\right\}
\ea
where $\phi$ denotes the azimuthal angle between the lepton and the hadron plane and 
the angle $\theta$ defines the direction of the negatively charged lepton momentum
in the rest frame of the virtual photon. The four partial cross sections are 
analogously defined to those for pion leptoproduction. Their evaluation within the
handbag approach is also analogous to pion production discussed in Sect.\ 1. The
only new item in this evaluation is the time-like electromagnetic form factor of the pion
replacing the space-like one. For $Q^{\prime 2}\geq 2\,\gev$ the average value
of the data from CLEO \ci{cleo}, BaBar \ci{babar} and from the $J/\Psi\to \pi^+\pi^-$
decay \ci{PDG}   
\be
Q^{\prime 2} |F_\pi(Q^{\prime 2})|\,=\, 0.88\pm 0.04\,\gev^2\,.
\label{eq:pionFF}
\ee
For its phase, $\exp{[i\delta(Q^{\prime 2})]}$, it is relied on a recent dispersion 
analysis \ci{belicka} which provides
\be
\delta \,=\, 1.014 \pi + 0.195(Q^{\prime 2}/\hspace*{-0.005\tw}\gev^2 -2)  
          - 0.029(Q^{\prime 2}/\hspace*{-0.005\tw}\gev^2 -2)^2\,.
\ee
for $2\,\gev^2 \lsim Q^{\prime 2} \lsim 5\,\gev^2$.
In the absence of any other information on this phase we use this parametrization
up to $\approx 8.9\,\gev^2$ where $\delta=\pi$. For larger values of $Q^{\prime 2}$
we take $\delta=\pi$, the asymptotic phase of the time-like pion form factor 
obtained by analytic continuation of the perturbative result for the space-like 
form factor.  

\begin{figure}[th]
\centering
\includegraphics[width=0.32\tw]{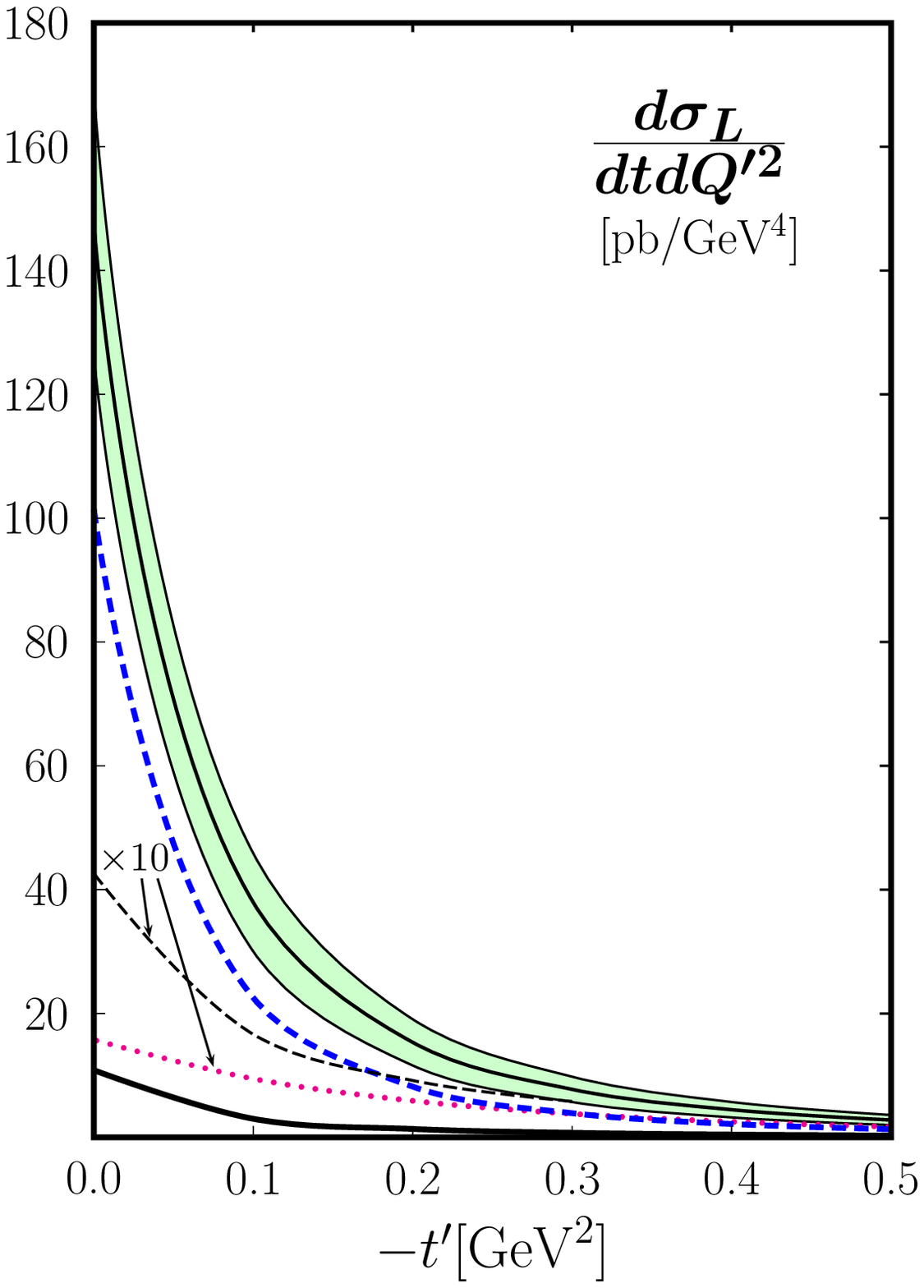}
\hspace*{0.06\tw}
\includegraphics[width=0.315\tw]{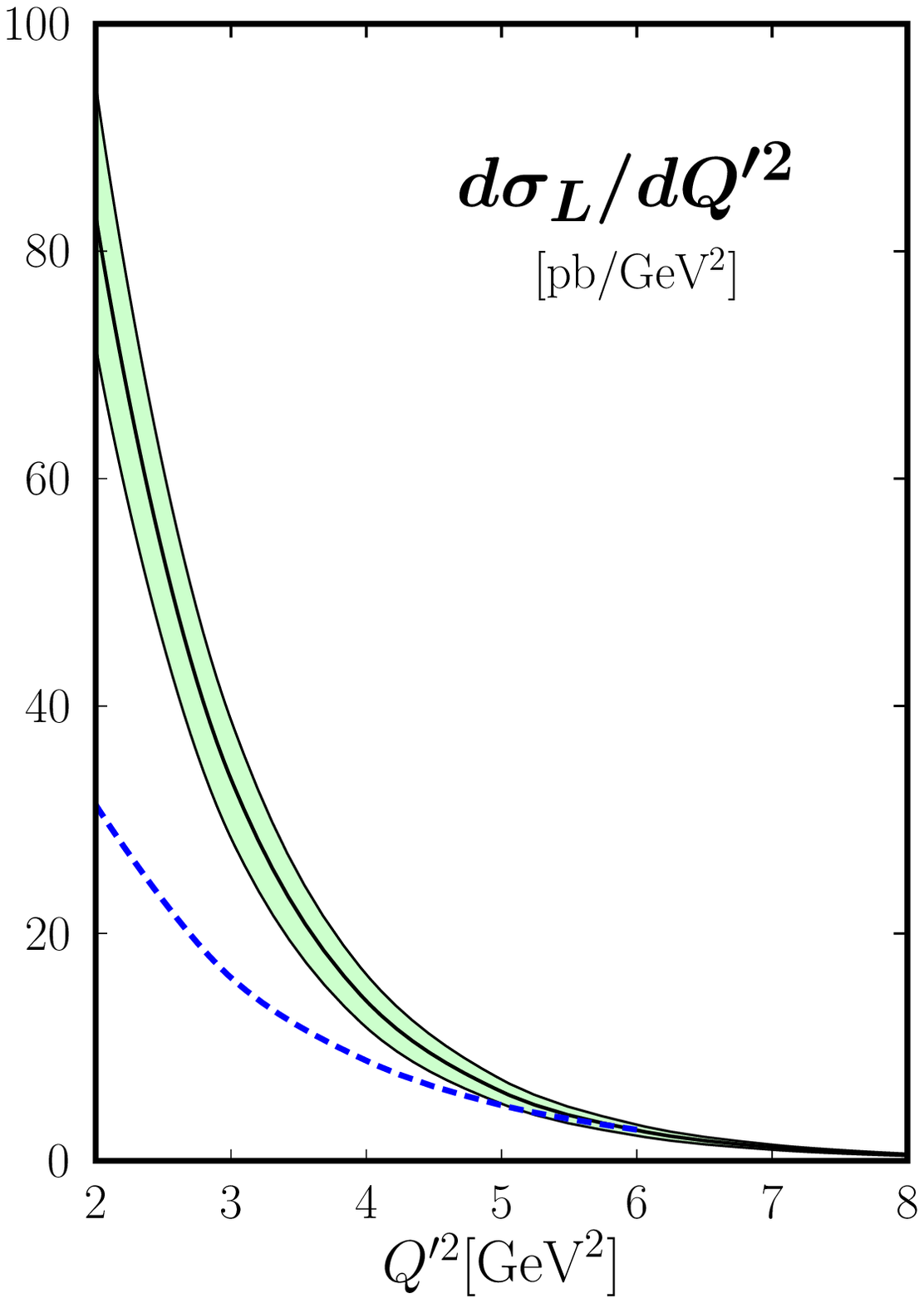}
\caption{The longitudinal cross sections $d\sigma_L/dtdQ^{\prime 2}$ (left) at 
$Q^{\prime 2}=4\,\gev^2$ versus $t'$  and $d\sigma_L/dQ^{\prime 2}$ (right) versus 
$Q^{\prime 2}$. The thin solid lines with error bands represent our full results 
at $s=20\,\gev^2$, the thick dashed ones those at $30\,\gev^2$. The thick solid
(dotted, thin dashed) line is the interference term (contribution from 
$|\langle \widetilde{H}^{(3)}\rangle|^2$, leading twist). The latter two results 
are multiplied by 10 for the ease of legibility.}
\label{fig:sigmaL}
\end{figure}

In Fig.\ \ref{fig:sigmaL} predictions for the longitudinal cross sections
are shown. They are markedly larger than the leading-twist results \ci{berger,sawada}.
This amplification is due to the use of the experimental 
value of the pion form factor \req{eq:pionFF} instead of its leading-twist result 
($\approx 0.15\,\gev^2$). We stress that the OPE contribution from the pion pole 
does neither rely on QCD factorization nor on a hard scattering. It is therefore 
not subject to evolution and higher-order perturbative QCD corrections. In \ci{GK9} predictions 
for the other partial cross sections are also presented. The transverse cross section,
which amounts to about $25\%$ of the longitudinal cross section, is dominated by
the GPD $H_T$. The contribution from $\bar{E}_T$ is tiny and, hence, $d\sigma_{TT}$.
On the other hand, the longitudinal-transverse interference cross section is not small. 
The cross sections decrease with growing $s$. At, say, $s\approx 360\,\gev^2$ as is available 
from the pion beam at CERN, the longitudinal cross section is about 
$30\,{\rm fb}/\gev^2$ at $Q^{\prime 2}=4\,\gev^2$. 

\section{Summary}
\label{summary}
In this article it is reported on calculations of $lp\to l\pi^+n$ and
$\pi^-p\to l^+l^-n$ within the handbag approach. Forced by experimental
results on hard pion production two corrections to the leading-twist amplitudes
are taken into account: the pion pole is treated as a one-pion exchange 
and amplitudes for transverse photons are added. These amplitudes are
modeled by transversity GPDs in combination with a twist-3 pion wave function.
With this generalized handbag approach reasonable agreement with the data
on pion production is achieved. Future data on $\pi^-p\to l^+l^-n$, measured
for instance at J-PARC may allow for a test of factorization in the time-like
region. There is no proof for it but its validity seems to be plausible.
However, Qiu \ci{qiu} raised doubts on the factorization arguments for the
exclusive Drell-Yan process.

The Drell-Yan process offers an opportunity to check the dependence
of the $\pi\pi\gamma^*$ vertex on the pion virtuality by comparing data
on the time-like pion form factor measured in $l^-l^+\to \pi^-\pi^+$,
with parametrizations of $\pi^-\pi^{*+}\to l^-l^+$ as part of the
Drell-Yan amplitudes. The extraction of the space-like form factor from
$lp\to l\pi^+ n$ may benefit from that check. 



\begin{thebibliography}{9}

\bibitem{collins96} J.~C.~Collins, L.~Frankfurt and M.~Strikman,
  Phys.\ Rev.\ D {\bf 56}, 2982 (1997).

\bibitem{GK9}S.~V.~Goloskokov and P.~Kroll,
  Phys.\ Lett.\ B {\bf 748}, 323 (2015).


\bibitem{hermes-aut} A.~Airapetian {\it et al.} [HERMES Collaboration],
  Phys.\ Lett.\ B {\bf 682}, 345 (2010).

\bibitem{clas-pi0} I.~Bedlinskiy {\it et al.} [CLAS Collaboration],
  Phys.\ Rev.\ C {\bf 90}, no. 2, 025205 (2014)
  [Phys.\ Rev.\ C {\bf 90}, no. 3, 039901 (2014)].

\bibitem{GK5} S.~V.~Goloskokov and P.~Kroll,
  Eur.\ Phys.\ J.\ C {\bf 65}, 137 (2010).

\bibitem{GK6} S.~V.~Goloskokov and P.~Kroll,
  Eur.\ Phys.\ J.\ A {\bf 47}, 112 (2011).

\bibitem{liuti15} G.~R.~Goldstein, J.~O.~Gonzalez Hernandez and S.~Liuti,
  Phys.\ Rev.\ D {\bf 91}, no. 11, 114013 (2015)

\bibitem{braun90} V.~M.~Braun and I.~E.~Filyanov,
  Z.\ Phys.\ C {\bf 48}, 239 (1990)
  [Sov.\ J.\ Nucl.\ Phys.\  {\bf 52}, 126 (1990)]
  [Yad.\ Fiz.\  {\bf 52}, 199 (1990)].

\bibitem{GK3} S.~V.~Goloskokov and P.~Kroll,
  Eur.\ Phys.\ J.\ C {\bf 53}, 367 (2008).


\bibitem{li-sterman} H.~n.~Li and G.~F.~Sterman,
  Nucl.\ Phys.\ B {\bf 381}, 129 (1992).

\bibitem{mankiewicz99} L.~Mankiewicz, G.~Piller and A.~Radyushkin,
  Eur.\ Phys.\ J.\ C {\bf 10}, 307 (1999).

\bibitem{F-pi-08} H.~P.~Blok {\it et al.} [Jefferson Lab Collaboration],
  Phys.\ Rev.\ C {\bf 78}, 045202 (2008).

\bibitem{hermes08}A.~Airapetian {\it et al.} [HERMES Collaboration],
  Phys.\ Lett.\ B {\bf 659}, 486 (2008).

\bibitem{FGHK15} L.~Favart, M.~Guidal, T.~Horn and P.~Kroll,
  Eur.\ Phys.\ J.\ A {\bf 52}, no. 6, 158 (2016).

\bibitem{GK8} S.~V.~Goloskokov and P.~Kroll,
  Eur.\ Phys.\ J.\ A {\bf 50}, no. 9, 146 (2014).

\bibitem{musatov00} I.~V.~Musatov and A.~V.~Radyushkin,
  Phys.\ Rev.\ D {\bf 61}, 074027 (2000).

\bibitem{DK13} M.~Diehl and P.~Kroll,
  Eur.\ Phys.\ J.\ C {\bf 73}, no. 4, 2397 (2013).

\bibitem{anselmino08} M.~Anselmino, M.~Boglione, U.~D'Alesio, A.~Kotzinian, 
  F.~Murgia, A.~Prokudin and S.~Melis,
  Nucl.\ Phys.\ Proc.\ Suppl.\  {\bf 191}, 98 (2009).


\bibitem{gockeler05} M.~Gockeler {\it et al.} [QCDSF and UKQCD Collaborations],
  Phys.\ Lett.\ B {\bf 627}, 113 (2005).

\bibitem{gockeler07} M.~Gockeler {\it et al.} [QCDSF and UKQCD Collaborations],
  Phys.\ Rev.\ Lett.\  {\bf 98}, 222001 (2007).

\bibitem{weiss16} P.~Schweitzer and C.~Weiss,
  PoS QCDEV {\bf 2015}, 041 (2015).

\bibitem{hall-A} M.~Defurne {\it et al.},
  arXiv:1608.01003 [hep-ex].

\bibitem{kubarowsky10} V.~Kubarovsky [for the CLAS Collaboration],
  Int.\ J.\ Mod.\ Phys.\ Conf.\ Ser.\  {\bf 40}, 1660051 (2016).

\bibitem{berger}  E.~R.~Berger, M.~Diehl and B.~Pire,
  Phys.\ Lett.\ B {\bf 523}, 265 (2001).

\bibitem{sawada} T.~Sawada, W.~C.~Chang, S.~Kumano, J.~C.~Peng, S.~Sawada and K.~Tanaka,
  Phys.\ Rev.\ D {\bf 93}, no. 11, 114034 (2016).


\bibitem{cleo} K.~K.~Seth, S.~Dobbs, Z.~Metreveli, A.~Tomaradze, T.~Xiao 
         and G.~Bonvicini,
  Phys.\ Rev.\ Lett.\  {\bf 110}, no. 2, 022002 (2013).

\bibitem{babar} B.~Aubert {\it et al.}  [BaBar Coll.],
  Phys.\ Rev.\ Lett.\  {\bf 103}, 231801 (2009).

\bibitem{PDG} K.~A.~Olive {\it et al.} [Particle Data Group Collaboration],
  Chin.\ Phys.\ C {\bf 38}, 090001 (2014).

\bibitem{belicka} M.~Belicka, S.~Dubnicka, A.~Z.~Dubnickova and A.~Liptaj,
  Phys.\ Rev.\ C {\bf 83}, 028201 (2011).


\bibitem{qiu} J. Qiu, talk presented at the KEK workshop on 'Hadron physics
with high-momentum hadron beams at J-PARC', March 2015.
\end{thebibliography}
\end{document}